\documentclass{aa}

\usepackage{graphicx}
\usepackage{amsmath,amssymb}
\usepackage{txfonts,natbib}

\def\grays{$\gamma$-rays}
\def\gray{$\gamma$-ray}

\newcommand{\jetp}{{JETP}}
\newcommand\sax{{BeppoSAX}}

\begin{document}

  \title{Non-thermal bremsstrahlung from supernova remnants and the effect of 
Coulomb losses}


   \author{Jacco Vink
      \inst{1}
          }

   \offprints{J. Vink (j.vink@astro.uu.nl)}

 \institute{Astronomical Institute Utrecht, Utrecht University, 
P.O. Box 80000 3508TA, Utrecht, The Netherlands\\  \email{j.vink@astro.uu.nl}
}

\abstract{}
{I investigate the shape of the electron cosmic ray spectrum
in the range up to $\sim 1000$~keV, assuming that the acceleration process
at the shock 
results in a power law in momentum, and that downstream of the
shock the spectrum
is affected by Coulomb interactions with background electrons only.
}{
In the non-relativistic regime one can analytically determine the
energy of an electron starting with a certain energy, and use this
result to produce an electron cosmic ray spectrum, modified by Coulomb losses.
}{
An analytic expression for the electron spectrum is obtained that depends
on the parameter $n_{\rm e}t$, which can be estimated from a
similar parameter used to characterize the line spectra of supernova remnants.
}{
For the brightest supernova remnants $n_{\rm e}t > 10^{11}$~cm$^{-3}$s,
and most of the electrons accelerated to $< 100$~keV have lost
their energy. Because of its high radio flux,
Cas A is the most likely candidate for non-thermal bremsstrahlung.
Although it has $n_{\rm e}t \sim 2\times 10^{11}$~cm$^{-3}$s, one
may expect to pick up non-thermal bremsstrahlung above 100~keV with current
hard X-ray detectors.
}

 \maketitle

\section{Introduction}
Over the last five years there has been considerable progress in
our understanding of high energy cosmic ray acceleration by supernova
remnants (SNRs).
The spatial distribution and spectral shape of 
X-ray synchrotron emission from all historical SNRs and some
other SNRs has made it clear that electron can be accelerated to
energies up to $\sim 100$~TeV \citep{koyama95}, 
that diffusive shock acceleration works very efficiently, with diffusion
close to the Bohm limit \citep{vink05b,vink06d,parizot06,stage06}, and that
magnetic fields are amplified by cosmic ray streaming 
\citep{vink03a,berezhko03c,bamba05,voelk05}. Moreover, the
morphology of Tycho's SNR shows that at least in this remnant, but
also probably in Kepler's SNR, cosmic ray acceleration is so efficient that
it gives rise to enhanced compression ratios \citep{warren05,ellison04}.
In addition, considerable progress has been made in the field of
TeV astronomy, with many SNRs being established as sources
of TeV \grays\ \citep[e.g.][]{aharonian04,aharonian05,albert07}.
However, it has not yet been firmly established whether the \gray\
emission is dominated by inverse Compton scattering from TeV electrons,
or from neutral pion decay, caused by TeV ion cosmic rays.

In contrast,
little has happened concerning observations of low energy cosmic rays. 
This is unfortunate, since from a theoretical perspective, the initial
stages of the acceleration process, from thermal energies up to energies where
Fermi acceleration efficiently operates, is complex
and not well understood \citep[see e.g.][for a review]{malkov01}.
In fact, the processes by which particles are injected
into the Fermi acceleration process, are likely to be 
different for electrons
and ions, unlike Fermi acceleration itself. 
Our current understanding of
the initial stages of acceleration
is largely based on either in situ observations of
inter-planetary shocks,
and on computer modeling using hybrid or particle in cell codes
\citep[e.g.][]{bykov99,schmitz02,lee04}.
There was some hope that hard X-ray observations of SNRs
might provide  observational information on at least the electron component
of low energy cosmic rays \citep{asvarov90,favata97,vink97,vink03a}. 
However,
it is generally agreed that the non-thermal hard X-ray emission 
from SNRs such as Cas A \citep{the96,allen97,favata97,vink03a,renaud06},
SN1006 \citep{allen99,kalemci06b}, Tycho \citep{allen99}, 
Kepler and RCW 86 \citep{allen99}
is due to X-ray synchrotron from the highest energy electron cosmic rays,
rather than from bremsstrahlung from low energy electrons.
Nevertheless, the issues remains of interest as current and future
hard X-ray detectors push the detection of hard X-ray emission 
to higher photon energies. 
At high enough energies, the steepening 
X-ray synchrotron spectrum is
likely to be overtaken by non-thermal bremsstrahlung.

In this paper I discuss the shape of the low energy
cosmic ray electron spectra,  given the importance of Coulomb-losses.
These Coulomb losses in general alter the shape of the low electron cosmic
ray spectrum as generated near the shock front. Note that a
similar situation was investigated by \citet{sarazin99} 
for higher energy electrons in clusters of galaxies.

\begin{figure}
\centerline{
\includegraphics[angle=-90,width=0.45\textwidth]{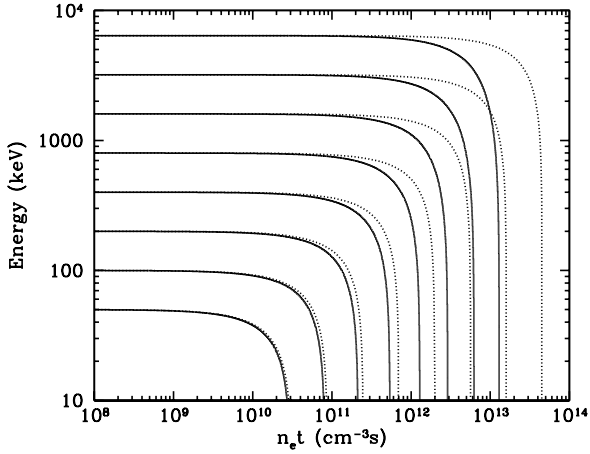}
}
\caption{
The electron energy as a function of $n_{\rm e} t$ 
($\lambda_{\rm ee} = 30.9$) for different values of energy at $t=0$~s
(50~keV, 100~keV, 200~keV...6400~keV).
The solid lines show the exact, but implicit solutions to
Eq.~\ref{eq-losses}, i.e. Eq.~\ref{eq-implicit}.
The dotted lines are the non-relativistic solutions to Eq.~\ref{eq-losses}, 
i.e. Eq.~\ref{eq-e_t}.
\label{fig-eltrajectories}
}
\end{figure}

\section{Coulomb loss affected electron spectra}
\subsection{The electron energy as a function of time}
Supra-thermal electrons loose energy through various processes:
bremsstrahlung losses, Coulomb losses (collisions with electrons/ions),
and ionization losses.\footnote{{ 
In contrast, in the GeV-TeV range the dominant losses are due to
synchrotron radiation and Coulomb losses, limiting the maximum
electron energy that can be obtained \citep[e.g.][]{drury99}.} }
In the hot plasmas inside SNR shells Coulomb
losses are likely to be the dominant source of energy losses of
electrons, in particular through electron-electron collisions 
\citep[e.g.][]{haug04}. Electron-electron collisions are most efficient
at low electron energies. For SNRs we are interested in the bremsstrahlung
emission of supra-thermal electron with energies in the range of 
$\sim 10-1000$~keV. The  problem with this energy range is that neither
the non-relativistic, nor the relativistic approximations for Coulomb
losses are completely valid. However, I will show that for most SNRs the 
non-relativistic approximation is sufficient.

The energy loss rate for an electron with $E >> kT_{\rm e}$ is given by 
\citep[e.g.][]{haug04,nrlplasma}:
\begin{equation}
  \frac{dE}{dt} = -4\pi r_{\rm e}^2 m_{\rm e} c^3\frac{\epsilon + 1}{p} 
  \lambda_{\rm ee} n_{\rm e},
  \label{eq-losses}
\end{equation}
with $r_{\rm e}$ the classical electron radius, 
and $\epsilon$ and $p$ the kinetic energy and momentum in dimensionless units,
$n_{\rm e}$ the electron density, and $\lambda_{\rm ee}$ the Coulomb logarithm,
given by \citep{nrlplasma}:
\begin{equation}
  \lambda_{\rm ee} = 30.9 - 
  \ln\Bigl[ n_{\rm e}^{1/2} \Bigl(\frac{1 {\rm keV}}{kT_{\rm e}} \Bigr) \Bigr]
\end{equation}
For  the non-relativistic limit ($\epsilon << 1$) Eq.~\ref{eq-losses} gives:
\begin{equation}
  \frac{dE}{dt} = 
-\frac{4\pi r_{\rm e}^2 m^2c^4}{\sqrt{2 m_{\rm e} E}}  
  \lambda_{\rm ee} n_{\rm e} = 
-\frac{ 7.73\times 10^{-6}}{\sqrt{E}}  \lambda_{\rm ee} n_{\rm e},
\label{eq-losses-nr}
\end{equation}
with energies, $E$ in units of eV.

The solution to this differential equation is:
\begin{equation}
E(t) = 
\bigl( E_0^{3/2} - 1.16\times 10^{-5} \lambda_{\rm ee}n_{\rm e} t\bigr)^{2/3},
\label{eq-e_t}
\end{equation}
with $E_0$ the energy of the electron at $t=0$.

In fact, the correct equation, Eq.~\ref{eq-losses}, 
can also be solved, but only gives
an implicit function for $E(t)$:
\begin{align}
n_{\rm e} t = 
\frac{  
\sqrt{ (\frac{E_0}{mc^2} + 1)^2 - 1} - 
\arctan\sqrt{ (\frac{E_0}{mc^2} + 1)^2 - 1 }\label{eq-implicit}
}
{4\pi r_{\rm e}^2 c \lambda_{\rm ee}} - \\ \nonumber
\frac{  
\sqrt{ (\frac{E}{mc^2} + 1)^2 - 1} - 
\arctan \sqrt{ (\frac{E}{mc^2} + 1)^2 - 1 }
}
{4\pi r_{\rm e}^2 c \lambda_{\rm ee}}.
\end{align}

Figure~\ref{fig-eltrajectories} shows both the full solution
and the non-relativistic approximation. As expected, the
non-relativistic approximation fails at high energies. 
For energies $\lesssim 400$~keV the approximation is reasonable,
with errors in the loss timescale of less than 30\%.

\subsection{An analytic approximation to the non-thermal 
spectrum}
The advantage of an explicit function (Eq.~\ref{eq-e_t}) is that
it can be easily used to calculate the evolution of the electron spectrum,
using the transformation  $N(E_0) \rightarrow N(E(t))$,
if an analytic approximation is known. { This can be done
by inverting Eq.~\ref{eq-e_t} and insert the result in $N(E)$,
taking into account that}
\begin{equation}
\frac{dE_0}{dE} = (E^{3/2} + 1.16\times 10^{-5} \lambda_{\rm ee} n_{\rm e} t)^{-1/3} E^{1/2}.\label{eq-dEdE0}
\end{equation}

A simple example is when $N(E)$ can be approximated by a power law 
$N(E_0)dE_0 = K E_0^{-\Gamma} dE_0$. After some time $t$ the spectrum
will be approximated by:
\begin{equation}
N(E) dE = K (E^{3/2} + 1.16\times 10^{-5} \lambda_{\rm ee} n_{\rm e}t)^{-(2\Gamma+1)/3} E^{1/2}dE. \label{eq-powl}
\end{equation}

\begin{figure}
\centerline{
\includegraphics[angle=-90,width=0.45\textwidth]{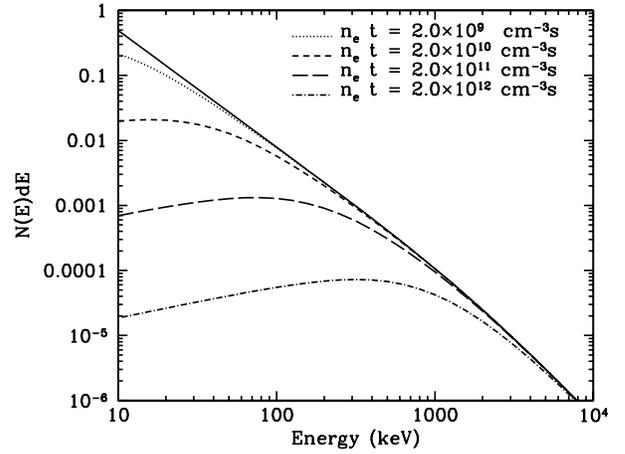}
}
\caption{
Low energy electron cosmic rays as expected from first order
Fermi acceleration, 
including the effects of Coulomb losses (Eq.~\ref{eq-pl_mom_nr}). The solid line gives the expected spectrum near the shock front,
whereas the other lines show the models for different values of
$n_{\rm e}t$.
\label{fig-elspectra}}
\end{figure}

A more interesting function in this context is the power law in momentum
$N(p) = K p^{-\Gamma}$, since the simplest version of first order
Fermi acceleration predicts that this is the spectrum of cosmic rays.
Changing this into a function of energy 
($E = \sqrt{ p^2c^2 + (mc^2)^2} - mc^2$)
gives \citep[see][]{asvarov90}:
\begin{equation}
N(E) dE  =  \frac{K}{c^{-\Gamma}} (E^2 + 2Emc^2)^{-\frac{\Gamma+1}{2}} \times 
 (E+mc^2) dE. \label{eq-pl_mom_energy}
\end{equation}
Inserting Eq.~\ref{eq-e_t} and \ref{eq-dEdE0} in this expression gives:
\begin{align}
\label{eq-pl_mom_nr}
 N(E)dE   & = \frac{N_p}{c^{-\Gamma}} \times & \\ \nonumber
 &\Bigl( (E^{3/2} +  1.16\times 10^{-5} \lambda_{\rm ee} n_{\rm e} t)^{4/3} + \\ \nonumber
& \hskip 1cm 2mc^2 (E^{3/2} +  1.16\times 10^{-5} \lambda_{\rm ee} n_{\rm e} t)^{2/3} \Bigr)^{-(\Gamma+1)/2} \times \\\nonumber
& \Bigl( (E^{3/2} +  1.16\times 10^{-5} \lambda_{\rm ee} n_{\rm e} t)^{2/3} + mc^2 \Bigr) \times \\ \nonumber
&\Bigl( E^{3/2} +  1.16\times 10^{-5} \lambda_{\rm ee} n_{\rm e} t \Bigr)^{-1/3} E^{1/2} dE.
\end{align}

One of the important aspects of this solution is that it depends almost
entirely on one parameter $n_{\rm e}t$, since $ \lambda_{\rm ee}$ only
weakly depends on the plasma density and temperature. This parameter
is familiar to anyone investigating X-ray emission from SNRs, as
it is the key parameter for non-ionization equilibrium 
\citep[NEI, see][]{liedahl99} and for the equilibration of electrons and ions.
In both cases the parameter is a measure for 
the number of collisions with
background electrons.

Fig.~\ref{fig-elspectra} shows the electron distributions for
values of  $n_{\rm e} t$ covering the relevant range for SNRs,
$10^9-10^{12}$~cm$^{-3}$s.
Fig.~\ref{fig-eltrajectories} shows that for young SNRs
the non-relativistic approximation is expected to be valid, as large
errors only appear for $n_{\rm e}t \gtrsim 5\times 10^{11}$~cm$^{-3}$s.
Moreover, the errors mostly affect the spectral shapes above 500~keV.

\begin{figure*}
\centerline{
\includegraphics[angle=-90,width=0.45\textwidth]{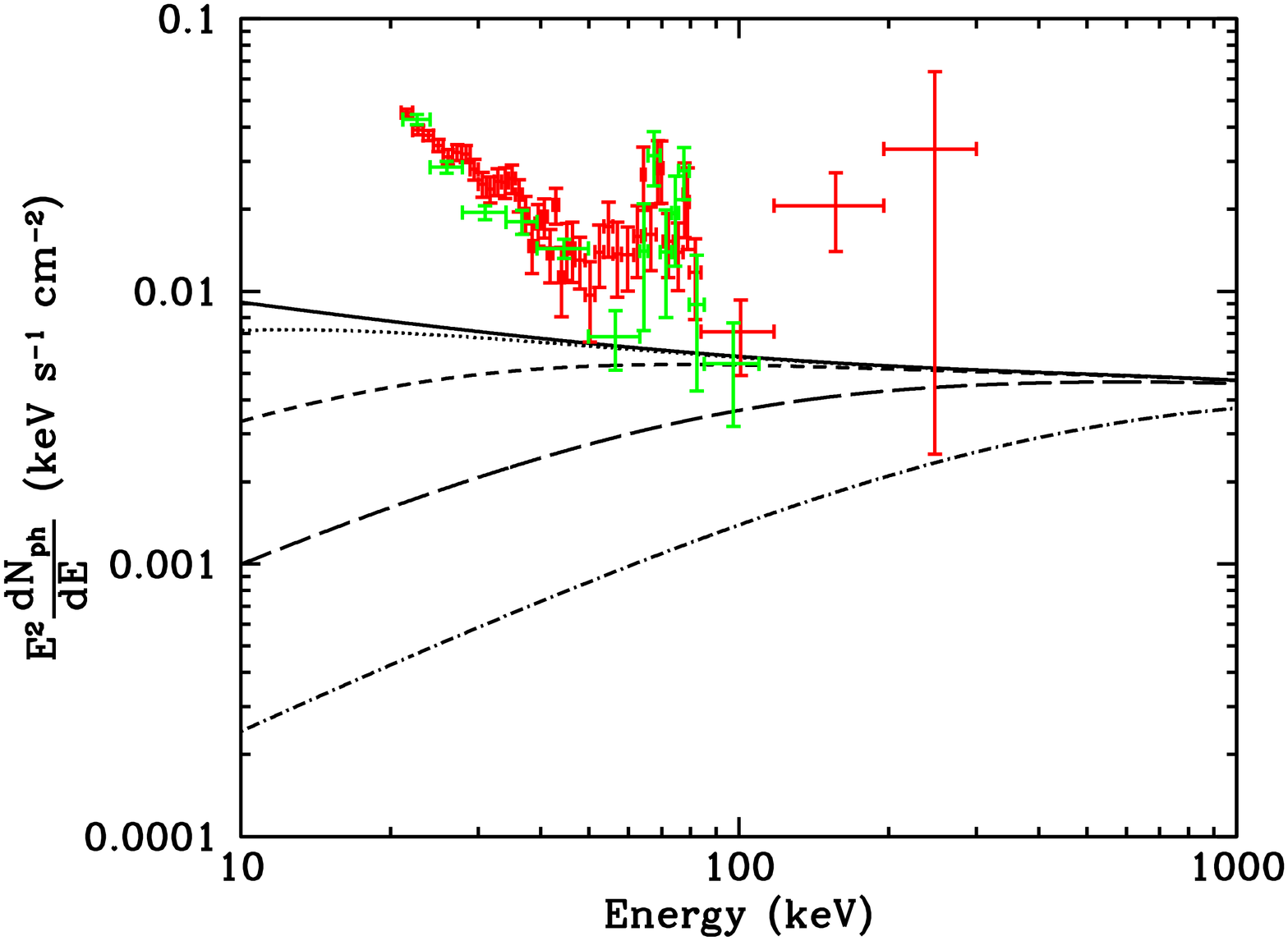}
\includegraphics[angle=-90,width=0.45\textwidth]{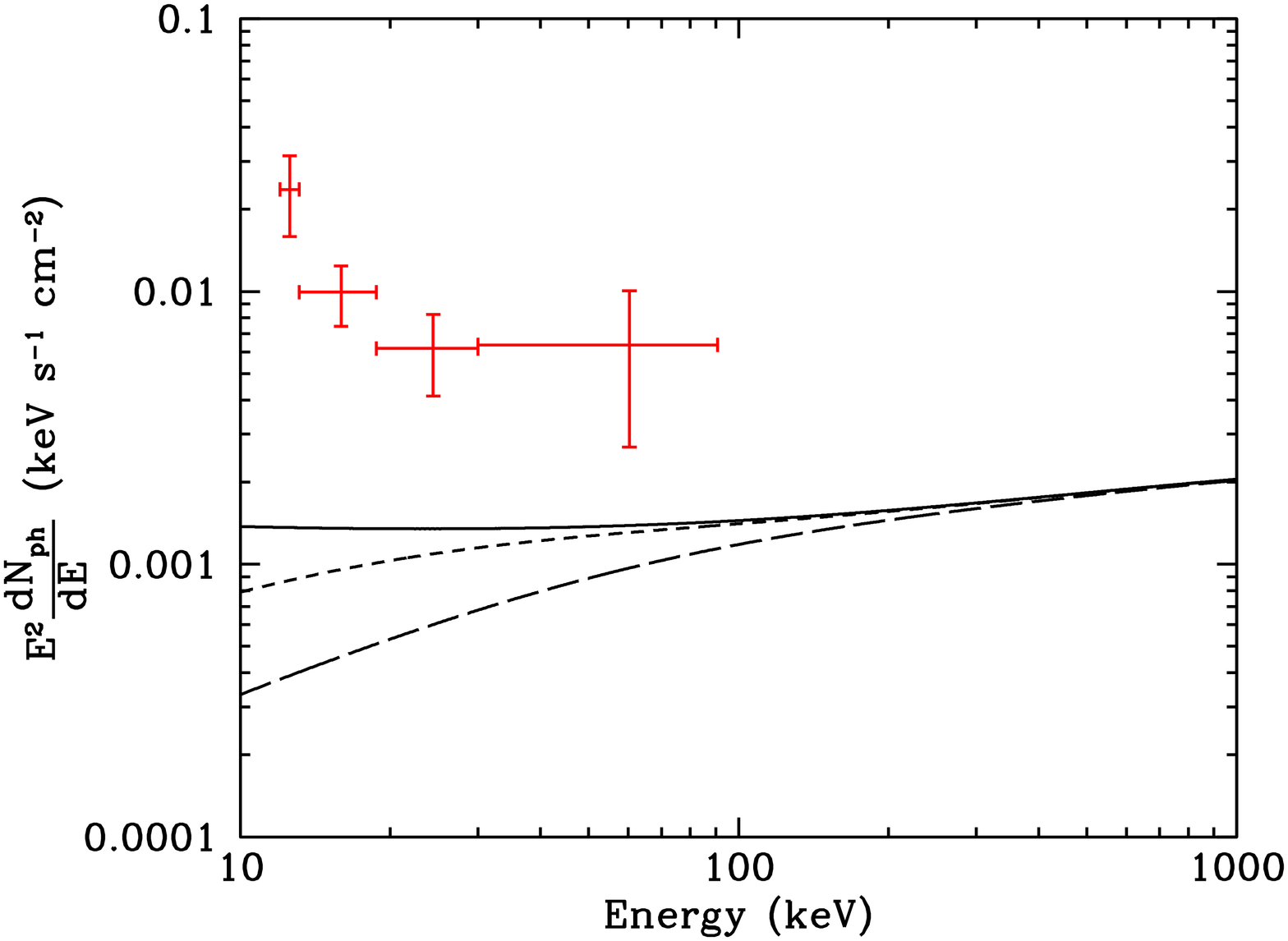}
}
\caption{
Left:
Spectral energy distribution of Cas A, as measured with 
\sax-PDS \citep{vink01a} and INTEGRAL-IBIS \citep{renaud06},
assuming a magnetic field of $\sim 300~\mu$G, a spectral energy index
of -2.56 (corresponding to a radio-spectral index of $\alpha=-0.78$),
and a { background plasma with $\Sigma_i <n_i Z^2>=10$~cm$^{-3}$.
The bremsstrahlung spectra were calculated 
using the analytic cross sections of \citet{haug97}.}
The model spectra are shown for  
$n_{\rm e}t = 0, 2\times 10^9,2\times 10^{10}, 2\times 10^{11}$  and
$2\times 10^{12}$~cm$^{-3}$s (from top to bottom).
Right: Spectral energy distribution of Tycho's SNR, as measured by the
\sax-PDS.
The theoretical curves are for $B=10~\mu$G, $\alpha=-0.6$, 
and $\Sigma_i <n_i Z^2>=4$~cm$^{-3}$. The different curves are
for $n_{\rm e}t = 0, 1\times 10^{10}, 1\times10^{11}$~cm$^{-3}$s.
\label{fig_casa}}
\end{figure*}

\section{Discussion and application to Cas A}

As explained above the shape of the non-thermal electron spectrum below
1000~keV depends on the parameter $n_{\rm e}t$, which is the same
parameter that governs the ionization state of the SNR. However,
it depends on the details of the shock acceleration history, whether one can
safely assume that the average $n_{\rm e}t$ of the hot plasma corresponds
to the  $n_{\rm e}t$ governing the shape of the low energy electron cosmic
rays. A close correspondence is likely if 
shock heating results always in a fixed ratio of heated and accelerated
electrons at the shock front. { Realistically, one should
take into account the variation of $n_{\rm e}t$ throughout the SNR, and
the evolution with time. However, in practise 
a one or two values for $n_{\rm e}t$
characterizes the overall plasma characteristics reasonably well 
\citep[e.g.][]{vink96}.}

For young SNRs like Cas A, Tycho, Kepler and SN1006, the
X-ray spectra are dominated by line emission from ejecta, shock heated by
the reverse shock. This means that also  $n_{\rm e}t$ is determined
mostly by the plasma heated by the reverse shock, rather than by the
forward shock heated plasma.  
It is often assumed that the reverse shock has
a low magnetic field and therefore does not efficiently accelerate 
cosmic rays. However, \citet{helder08}
showed recently that the
reverse shock of Cas A is responsible for most of the X-ray synchrotron
emission, this implies that the reverse shock is capable of accelerating
particles as well.
This is a confirmation of recent theoretical calculations
by \citet{ellison05}. { Nevertheless, it is not quite clear what the
ratio is of the electrons accelerated by the forward shock, as opposed to 
the reverse shock. Moreover, the initial acceleration process for electrons
may be different: The reverse shocked plasma is
dominated by metals. 
In that case, a sizable fraction of the accelerated electrons 
may originate from ionization of (mildly) accelerated ions.}

{ Apart from the plasma $n_{\rm e}t$ there is another reason why
the $n_{\rm e}t$ for the  accelerated electrons is expected to be high 
in Cas A:}
Most of the electrons seem to
have been accelerated in the early life of the SNR. The best
evidence for this is the $\sim 1$\%/yr decline  in its radio flux,
which is best explained by adiabatic expansion of the relativistic electron
gas \citep{shklovsky68}. Cas A may be a special case since it is evolving
into a red super giant wind, which has a density falling of with radius as
$r^{-2}$. This means that the particle flux entering the forward shock at any
radius is more or less constant, whereas the shock velocity is dropping
with radius. As a result particle acceleration was more efficient early on,
suggesting a high  $n_{\rm e}t$.

Unfortunately, this also means that it is quite unlikely that the
non-thermal X-ray emission of Cas A below $\sim 100$~keV is dominated by 
non-thermal bremsstrahlung, either from the forward shock, or
from internal shocks, as assumed by \citep{laming01a, laming01b,vink03a}.
The reason is that all  the emission must then come from a limited fraction
of the total shock heated plasma for which $n_{\rm e}t < 10^{10}$~cm$^{-3}$s
{ \citep[see][for the spatial distribution of $n_{\rm e}t$ ]{willingale02,yang07}.}
If I simply assume a linear relation between age and $n_{\rm e}t$, 
less than 5\% of the plasma has $n_{\rm e}t < 10^{10}$~cm$^{-3}$s.
The lower hybrid wave model propagated by 
\citet{laming01a} remains an attractive
option for the injection of electrons into the Fermi acceleration process 
itself \citep[see also][]{ghavamian07}.

Nevertheless, Cas A remains one of the most important targets for
searching for non-thermal bremsstrahlung, but only above $\sim 100$~keV, where
the spectrum is less affected by Coulomb losses. The reason is
that Cas A is by far the brightest Galactic SNR in the radio.
This implies the ample presence of relativistic electrons, in combination
with a relatively high magnetic field of $100-500~\mu$G 
\citep{shklovsky68,vink03a,berezhko04a}. 
This is illustrated in
Fig.~\ref{fig_casa}, which suggests that with INTEGRAL-ISGRI we may be close
to a detection, or even that part of the flux density around
100~keV is caused by non-thermal bremsstrahlung \citep[c.f. the CGRO-OSSE
detection,][]{the96}. No deep Suzaku Hard X-ray Detector (HXD) 
\citep{takahashi07}
of Cas A exist, but a deep observation with this sensitive instrument, may
also be able to detect emission above 100~keV. 

{ 
The normalization of the 
spectrum in Fig.~\ref{fig_casa} is determined from
the radio flux, and assuming a mean magnetic field of $\sim 300~\mu$G.
The bremsstrahlung also depends on the emission measure:
${\rm EM} 
= \int \Sigma_i n_i n_{\rm e} Z_i^2 dV$, with $n_{\rm e}$ being,
here, the density of the accelerated electrons.
For Cas A the accelerated electrons are present in both
the forward shock heated plasma, with 
$n_{\rm H} \approx 10$~cm$^{-3}$,
and in the reverse shock heated plasma, which consists
of pure metals, probably dominated by almost completely ionized oxygen 
\citep[e.g.][]{vink96}. The emitting volume of the forward shock heated plasma
is $V_f \approx 1.2\times 10^{57}$~cm$^3$, based on a 
shock radius of 2.6 pc and a contact discontinuity at $\sim 2$~pc.
The reverse shock radius is $\sim 1.9$~pc \citep{helder08}, thus 
$V_r \approx 1.4\times 10^{56}$~cm$^3$.
Conservatively assuming that there is $\sim 1$~M$_{\odot}$\ of 
ionized oxygen \citep{vink96} and that the other metals contribute
little to the bremsstrahlung, one finds for the shocked ejecta
$\Sigma_i n_i Z_i^2 \approx 35$~cm$^{-3}$. For the volume averaged
normalization one finds
$(V_f n_{\rm H} +  V_r \Sigma_i  n_i Z_i^2)/(V_f + V_r) \approx 13$~cm$^{-3}$.
Note that both the bremsstrahlung and synchrotron luminosity
scale with the emitting volume.
It is not unreasonable to assume that the synchrotron emitting
volume is equal to the volume of the shock heated plasma.
However, it is uncertain how the accelerated electron density
and magnetic field is distributed over this volume.
So the normalization of
$\Sigma_i  <n_i Z_i^2> = 10$~cm$^{-3}$\ used in Fig.~\ref{fig_casa}
should be considered as a reasonable guess.
For a constant non-thermal bremsstrahlung emissivity a higher value
of $\Sigma  <n_i Z_i^2>$\ implies a higher magnetic field.
}

In contrast, for Tycho's SNR, due to its much lower radio luminosity, 
one only expects a detection of the non-thermal
bremsstrahlung, if the magnetic field turns out to be unexpectedly low: 
$B \lesssim 10~\mu$G. This is roughly the value
of the compressed average ISM magnetic field, whereas
estimates for the magnetic field in Tycho
indicates $B \sim 300~\mu$G \citep{voelk05}.

Finally, one could also consider SNRs in which the densities and
$n_{\rm e}t$ values are modest, and for which also the magnetic fields
are low; 
for example SN1006 \citep[$n_{\rm e}t = 2\times 10^9$~cm$^{-3}$s][]{vink03b} 
and the Northeast of RCW 86 
\citep[$n_{\rm e}t \sim 5\times 10^9$~cm$^{-3}$s][]{vink06d}.
The advantage is that although the bremsstrahlung may be weaker, the low
$n_{\rm e}t$ ensures non-thermal bremsstrahlung at lower photon energies.
Unfortunately, both these SNRs have also X-ray synchrotron radiation,
which may be hard to distinguish from non-thermal bremsstrahlung. However,
with imaging spectroscopy  above 10~keV one may isolate regions where
non-thermal bremsstrahlung is important and X-ray synchrotron radiation
absent. Moreover, high spectral resolution spectra
may reveal departures from Maxwellian electron distributions.
This type of research will have to wait till the emergence of
high resolution, high throughput,  X-ray imaging spectroscopy with
XEUS \citep{parmar06}.

\section{Summary and conclusions}

I have derived an analytic expression for the low energy electron
cosmic ray spectrum, assuming that the cosmic ray spectrum, as produced
by the diffusive acceleration process, 
is a power law in momentum, and is only affected by
Coulomb losses downstream of the shock. 
As may be expected, the parameter
governing the Coulomb losses depends on $n_{\rm e}t$, the product
of (background) electron density and the age of the accelerated
electron population. This parameter is similar to the ionization
time scale of the hot, shock heated plasma, and can be easily estimated
from the line spectra of SNRs.

For the SNRs with high
densities, and large $n_{\rm e}t$, one does not expect a large
population of accelerated electrons below 100~keV. This is in particular
the case for the very bright radio source Cas A. Nevertheless, Cas A 
remains a good candidate for searching for non-thermal bremsstrahlung,
because of its high ion density and because its radio brightness indicates
a large amount of accelerated electrons. 
Based on reasonable, but somewhat uncertain, assumptions, 
the results presented here
imply that only above $\sim 100$~keV one expects to pick up the non-thermal
bremsstrahlung component from Cas A. 
This may be accomplished with future deep Suzaku HXD
observations, or additional deep INTEGRAL observations.

\begin{acknowledgements}
I am grateful to Prof. Ishida and Dr Bamba for inviting me
for a short stay at ISAS, Japan in November 2007. The discussion on
Suzaku HXD data of Cas A led to the results reported here.
JV is financially supported by a Vidi grant from 
the Netherlands Organization for
Scientific Research (NWO).
\end{acknowledgements}

\end{document}